       \documentstyle[]{l-aa-dipastro}
       \input psfig.tex

	\newcommand{\Msun}{\mbox{${\rm M_{\odot}}$}}
	\newcommand{\Zsun}{\mbox{${\rm Z_{\odot}}$}}

        \newcommand{\Hbeta}{\mbox{${ H_{\beta}}$}}
        \newcommand{\MgFe}{\mbox{${ [MgFe]} $}}
        \newcommand{\MFe}{\mbox{${ \langle Fe \rangle}$}}
        \newcommand{\alfa}{\mbox{$\alpha$-elements}}

        \def\M12{${ M_{L,12}} $}
        \def\Mg2{${ Mg_{2}} $}

\def\smallskip{\vskip 8pt}
\def\littleskip{\vskip 6pt}

\begin{document}
\thesaurus{}

\title{To what Extent are \Mg2\ and \MFe\ Indicators
of Mg and Fe Abundances ? }

\author {R. Tantalo$^1$, A. Bressan$^2$, C. Chiosi$^{3,1}$}

\institute {
$^1$ Department of Astronomy, Vicolo dell' Osservatorio 5, 35122 Padua, 
   Italy\\
$^2$ Astronomical Observatory, Vicolo dell' Osservatorio 5, 35122 Padua, 
   Italy\\
$^3$ European Southern Observatory, Karl-Schwarzschild-Strasse 2,
   D-85748, Garching bei Muenchen, Germany}

\offprints{R. Tantalo }
\date {Received  October 1997; accepted}
\maketitle
\markboth{$Mg$ and $Fe$ abundances from line indices }{}

\begin{abstract}

The gradients in line strength indices such as \Mg2\ and \MFe\ (Carollo \&
Danziger 1994a,b; Carollo et al. 1993), and their increase toward massive
system (Worthey et al. 1992, Faber et al. 1992, Davies et al. 1993, Weiss et
al. 1995) observed in elliptical galaxies are customarily considered as
indicators of gradients in the chemical abundances of $Mg$ and $Fe$ and
arguments are given for an enhancement of $Mg$ ($\alpha$-elements in general)
with respect to $Fe$ toward the center of these galaxies or going from dwarf
to massive ellipticals. In this paper, we present a detailed analysis of the
whole problem aimed at understanding how the   indices \Mg2\ and \MFe\ depend
on the the chemical abundances and age of the stellar mix of a galaxy. We find
that  \Mg2\ and \MFe\ do not simply correlate with the abundances of $Mg$ and
$Fe$ and the ratio $Mg/Fe$, but depend also on the underlaying distribution of
stars in different metallicity bins $N(Z)$, or equivalently on the past
history of star formation. In practice, inferring the abundance of $Mg$ and
$Fe$ and their ratio is hampered by the need of some information about $N(Z)$.
Finally, the observational gradients in \Mg2\ and \MFe\ do not automatically
imply gradients in chemical abundances.

\end{abstract}

\section{Introduction}

The line strength indices \Mg2, \MFe, \Hbeta, etc. and their gradients are
customarily used to infer the  age and metallicity and their variations
across galaxies. 

Furthermore, in elliptical galaxies, the gradients in \Mg2\ and \MFe\ have
different slopes (Fisher et al. 1995,1996, Carollo \& Danziger 1994a,b;
Carollo et al. 1993, Davies et al. 1993), which is used as a clue to arguing that
$Mg$ (\alfa\ in general) is enhanced with respect to $Fe$ toward the center.

The bottom line to infer from \Mg2\ and \MFe\ an enhancement in \alfa\ rests
on the  notion that these two indices  depend on age  and the 
abundances of $Mg$ and $Fe$,  and that age and abundance 
effects can somehow be disentangled. If this is the case, the implications are
of paramount importance. It is worth recalling that according to the current
nucleosynthesis  scenario $Fe$ is mainly produced by Type Ia supernovae
(accreting white dwarfs in binary systems in the most popular scheme) and in
smaller  quantities by Type II supernovae. In contrast, only Type II supernovae
contribute to oxygen and \alfa. Furthermore as the mean lifetime of a binary
system (Type Ia progenitors) is $\geq 1$ Gyr, the contamination by Type I
supernovae occurs later as compared to  Type II supernovae. Finally, we expect
the iron abundance $[Fe/H]$ and the  $[\alpha/Fe]$ ratios  to increase and
decrease, respectively, as the galaxy ages.  In standard models of galactic
chemical evolution, i.e. constant initial mass function and supernova driven
galactic winds (cf. Matteucci 1997 for a comprehensive review of the subject),
this means that to obtain a galaxy (or region of it) enhanced in \alfa\ the
time scale of star formation over there must be shorter than about 1 Gyr. This
is a very demanding constraint on models of galaxy formation and evolution. It
is worth clarifying that such a conclusion is largely independent of the IMF
and galactic wind model in usage, even if some {\it ad hoc} combinations of IMF and/or galactic wind model can be found in which enhancement of \alfa\ is possible
irrespective of the argument about the time scale of star formation. The reader
is referred to the review article by Matteucci (1997) and the recent
study by Chiosi et al. (1997) on an unconventional IMF for more details.

In this paper, we address the question as to what extent the indices \Mg2\ and
\MFe\ (and their gradients)  depend on age and chemical abundances, paying
particular attention to the complex environment of a galaxy in which stars  of
many ages and chemical compositions are present. In other words, we seek to clarify how  the past history of star formation, which manifests itself in the extant   relative
distribution of stars per metallicity bin  [thereinafter the partition
function $N(Z)$], affects the correspondence between indices, ages and
abundances.

To this aim, we will utilize the spherical models of elliptical galaxies
developed by Tantalo et al. (1997),  which take into account gradients in mass
density, star formation rate, and chemical abundances (Section 2). With the
aid of these models, we predict how the gradients in $Mg$ and $Fe$ (and their
ratio) translate into gradients in \Mg2\ and \MFe, and  check whether a
gradient in \Mg2\ steeper than a  gradient in \MFe\ implies an enhancement of 
$Mg$ with respect to $Fe$ toward the center of these galaxies. We anticipate
here that, while these models are indeed able to match many  key properties of
elliptical galaxies, including the gradients in broad band colors (see below),
they lead to contradictory results as far as the gradients in line strength
indices are concerned (Section 3). To understand the physical cause of this
odd behaviour of the models, we check the calibration in use and the response
of \Mg2\ and \MFe\ to chemistry (Section 4). The reason of the contradiction
resides in the dependence of the indices in question on the existing $N(Z)$,
i.e. the past history of star formation. It will be shown that   gradients in
the indices \Mg2\ and \MFe\  do not automatically correspond to gradients in
the $Mg$ and $Fe$ abundances, and their ratios in particular (Section 4). In
order to cast more light on this topic, in a way independent of the particular
model of galactic evolution, we derive the above indices for single stellar
populations (SSP) of different metallicity and age, and look at the possible
combinations of these two parameters leading to the same values for the
indices (Section 5). Finally, some concluding remarks are drawn in Section
6.

\section{The reference model }

{\it Sketch of the model. }
Elliptical galaxies are supposed to be made of baryonic and dark matter both
with spherical distributions but different density profiles.  While dark matter
is assumed to have remained constant in time, the baryonic material (initially
in form of primeval gas) is supposed to have accreted at a suitable rate onto
the potential well of the former. The rate at which the density of baryonic
material  grows with time at any galacto-centric distance is chosen in such a
way that at the galaxy age $T_G$ it matches the radial density distribution
derived by Young (1976) for spherical systems. The density profile  of the
dark-matter is taken from  Bertin et al. (1992) and Saglia et al. (1992)
however adapted to the Young formalism for the sake of internal consistency.
The mass of dark matter is taken in fixed proportions with respect to the
luminous one, $ M_D=\theta \times M_L$ (all models below are for $\theta=5$), 
Dark matter will only affect the gravitational potential.

Given the total present day luminous mass $ M_L$ of the galaxy (thereinafter in
units of $ 10^{12}\times M_{\odot}$ and shortly indicated by $ M_{L,12}$) and
its effective radius $ R_{L,e}$, the model is divided into a number of
spherical shells, with proper spacing in radius and mass. The luminous mass of
each shell is written as

\begin{displaymath}
          \Delta M_{L,S} = \overline \rho_L(r) \times \Delta V(r), 
\end{displaymath}
where $\Delta V(r)$ is the volume of the shell and $\overline \rho_L(r)$ is
the mean density  of baryonic material, respectively, The inner and outer
radius of each shell are chosen in such a way that, using the Young (1976)
tabulations of the mean density as a function of the fractionary radius
$s=r/R_{L,e}$, the  mass $\Delta M_{L,S}$ amounts to about 5\% of the total
mass $M_{L,12}$.  The radial variation of the gravitational potential of dark
and luminous mass can be easily derived from the above description. The reader
is referred to Tantalo et al. (1997) for all other details.
\littleskip

{\it Accretion rate. }The collapsing rate of luminous material (gas)
is expressed as

\begin{equation}
\left[\frac{d{\rho_{L}}(r,t)}{dt}\right]_{inf} = 
            \rho_{L0}(r) \times exp({-\frac{t}{\tau(r)}})
\label{drho}
\end{equation}

\noindent
where $\tau(r)$ is the local time scale of gas accretion to be discussed
below, and  $\rho_{L0}(r)$ is fixed by imposing that at the present-day age of
the galaxy $T_{G}$ the density of luminous material in each shell has grown to
the value given by the Young profile.
\littleskip

{\it Star formation rate.} This follows the standard Schmidt law

\begin{equation}
\left[\frac{d{\rho_{g}}(r,t)}{dt}\right]_{sf} 
                    = \nu(r) {\overline \rho_{g}}(r,t)
\label{esfr}
\end{equation}

\noindent
where $\overline \rho_{g}(r,t)$ is the mean local gas density and $\nu(r)$ is
the specific efficiency of star formation.
\littleskip

{\it Accretion time scale $\tau(r)$. }
A successful description of the gas accretion phase is possible adapting to
galaxies the radial velocity law describing the final collapse of the core in
a massive star (Bethe 1990), i.e. free-fall [$v(r) \propto r^{-\frac{1}{2}}$]
in all regions external to a certain value of the radius $r^*$ and homologous collapse inside [$v(r) \propto r$]. This picture is also confirmed by Tree-SPH
dynamical models of elliptical galaxies (cf. Carraro et al. 1997 and references therein). Let us cast
the problem in a general fashion by expressing the velocity $v(r)$ as

\begin{displaymath}
       v(r) = c_1 \times r^{\alpha}  ~~~~~~~~~~ {\rm for}~~ r \leq ~~ r^*
\end{displaymath}
\begin{displaymath}
       v(r) = c_2 \times r^{- \beta} ~~~~~~~~ {\rm for}~~ r > ~~ r^*
\end{displaymath}

\noindent
(where $c_1$, $c_2$, $\alpha$ and $\beta$ are suitable constants), and the
time scale of accretion as

\begin{displaymath}
\tau(r)  \propto { r \over v(r) } 
\end{displaymath}

\noindent
In the models below we adopt  $\alpha=2$  (as suggested by the Tree-SPH calculations) and $\beta=0.5$ (as indicated by the core collapse analogy).
The determination
of the constants $c_1$ and $c_2$ is not strictly required as long as we seek
for scaling relationships. 

The time scale of gas accretion can be
written as proportional to some arbitrary time scale, modulated by a
correction term arising from the scaling law for the radial velocity. For the
time scale base-line we can take the free-fall time scale $t_{ff}$ referred to
the whole system,

\begin{equation}
\tau(r) = 
     t_{ff} \times \frac{r^{*}}{r} ~~~~~~~~~~~~
           { \rm if }~~ r \leq r^{*} 
\label{tau_ff_1}
\end{equation}

\begin{equation}
\tau(r) = 
     t_{ff} \times (\frac{r}{r^{*}})^{3/2} ~~~~~ 
                 { \rm if} ~~ r > r^{*} 
\label{tau_ff_2}
\end{equation}

\noindent
For the free-fall time scale $t_{ff}$ we make use of the relation by Arimoto
\& Yoshii (1987)

\begin{equation}
t_{ff} = 0.0697 \times  M_{L,12}^{0.325} ~~~~~~~~~~~{ Gyr}.
\label{tff}
\end{equation}
\noindent
Finally,  we take $r*= {1
\over 2} R_{L,e}$. Other choices are obviously possible without changing the
overall results of this study.
\littleskip

{\it Specific efficiency of star formation.}
In order to derive the specific efficiency of star formation $\nu(r)$ we
utilize the simple scale relations developed by Arimoto \& Yoshii (1987)
however adapted to the density formalism. At the typical galactic densities
($10^{-22}$ - $10^{-24} $ g~cm$^{-3}$) and considering hydrogen as the
dominant coolant (Silk 1977) the critical Jeans length is much smaller than
the galactic radius, therefore the galaxy gas can be considered as made of
many cloud lets whose radius is as large as the Jeans scale. If these clouds
collapse nearly isothermal without  suffering from mutual collisions, they will
proceed through subsequent fragmentation processes till opaque small subunits
(stars) will eventually be formed. In such a case the stars are formed on the
the free-fall time scale. In contrast, if mutual collisions occur,  they will
be supersonic giving origin to layers of highly cooled and compressed
material, the Jeans scale falls below the thickness of the compressed layer
and fragmentation occurs on the free-fall time scale of the high density
layers, and finally the whole star forming process is driven by the collision
time scale. On the basis of these considerations, we take the  ratio

\begin{equation}
 \sqrt{ \frac{1} {t_{ff} \times t_{col} } } 
\label{nu_star}
\end{equation}

\noindent
as a measure of the net efficiency of star formation.

\begin{figure}
\psfig{file=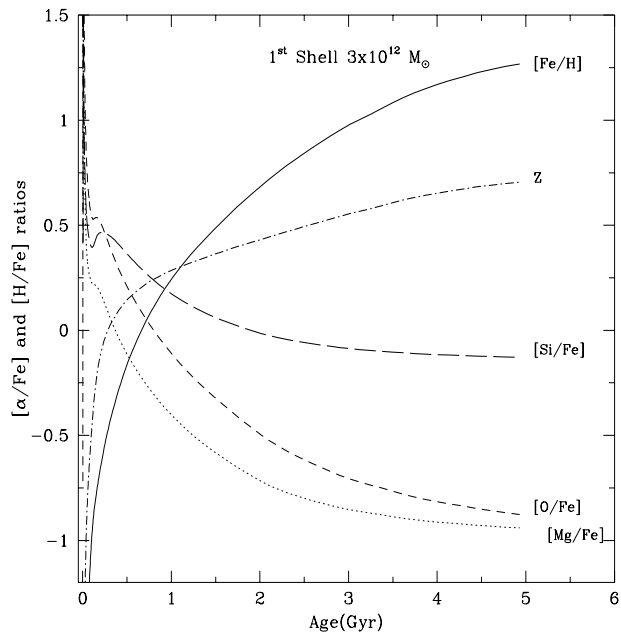,height=9.0truecm,width=8.5truecm} \caption{Temporal
evolution of four abundance ratios: $[Fe/H]$ ({\em solid line}), $[Mg/Fe]$
({\em dotted line}), $[O/Fe]$ ({\em dashed line}), and $[Si/Fe]$ ({\em
long-dashed line}). The {\em dot-dashed line} shows the ratio $Z/$\Zsun\ as a
function of time. The abundance ratios are in the standard notation}
\label{x_age}
\end{figure}

Let us express $\nu(r)$ as the product of a suitable yet arbitrary specific
efficiency $\nu^*$ referred to the whole galaxy and a dimensionless quantity
$F(r)$ describing as the ratio of eq. (\ref{nu_star}) varies with the
galacto-centric distance. An obvious expression for $F(r)$ is the ratio
(\ref{nu_star}) normalized to its central value.

According to Arimoto and Yoshii (1987) the mean collision time scale referred
to the whole galaxy can be expressed as

\begin{equation}
t_{col} = 0.0072  \times M_{L,12}^{0.1}   ~~~~~~~~~~~~~~Gyr
\label{t_jeans}
\end{equation}
\littleskip

With aid of this and the relation for the the free-fall time scale above
we can first calculate $\nu^*$

\begin{equation}
\nu^* = \left[ \sqrt{\frac{1}{t_{ff} \times t_{col}}} \right]_{gal} 
\label{nu_star_def}
\end{equation}

Extending by analogy the definition of free-fall and collision time scale
to each individual region, we get
 
\begin{equation}
F(r) = \left( { r_{c} \over r } \right) ^{3 \gamma }
          \times 
\left[\frac{ \overline{\rho}_{g}(r_{c},T_{G}) }
            {\overline{\rho}_{g}(r,T_{G}) }
            \right]^{\gamma} 
\label{fr}
\end{equation}

\noindent
where $\overline \rho_g(r,T_G)$ is the mean gas density within the region of
mid radius $r$ and  $r_c$  is the mid radius of the innermost sphere. 

\begin{figure*}[]
\psfig{file=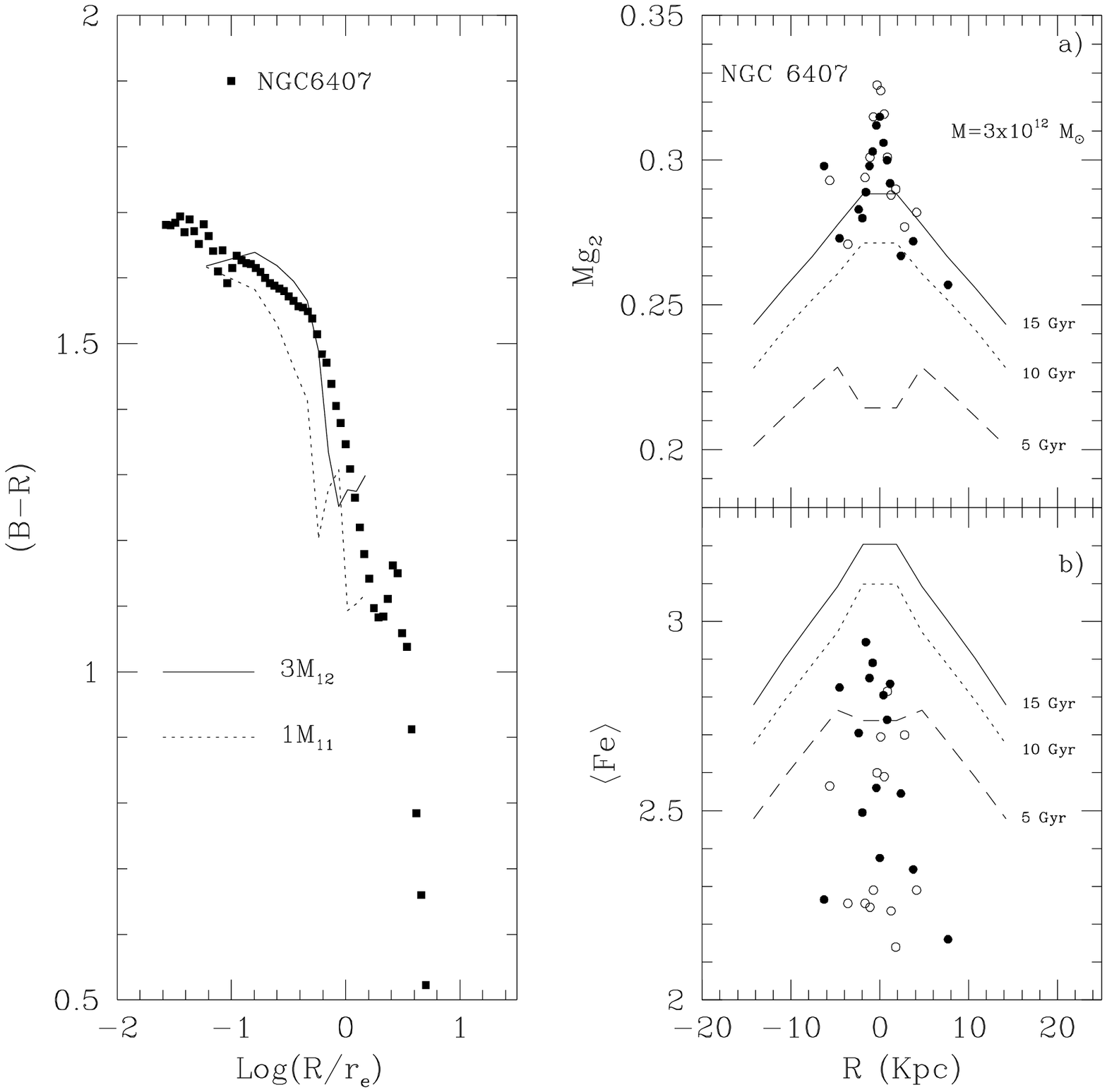,height=9.0truecm,width=17.0truecm}
\caption{Gradients in colors and line strength indices for the galaxy NGC~6407
(Carollo \& Danziger 1994a). {\it Left Panel}: comparison with the theoretical
gradients in (B--R) for models of different mass and same age (15 Gyr). {\it
Right Panel}: comparison with the theoretical gradients in line strength
indices for the $ 3 \times  M_{L,12}$ model which has nearly the same $
M/L_{B}$ ratio as NGC~6407. Finally, the observational data along major and
minor axis are indicated by  full and empty circles, respectively}
\label{mod_car}
\end{figure*}

In principle, the exponent $\gamma$ could be derived from combining the mass
dependence of $t_{ff}$ and $t_{col}$, i.e. $\gamma \simeq 0.2$. However, the many models calculated by Tantalo et al. (1997) show that in order to recover the
observational gradients in broad band colors 
 (and other properties of elliptical galaxies), the efficiency of star formation (i.e. $F(r)$ in our formulation) must vary with the radial distance more strongly than predicted by $\gamma=0.2$. 
The following relation is found to give good results

\begin{equation}
\gamma = 0.98\times  M_{L,12}^{0.02}
\label{alfa_nu}
\end{equation}

\noindent
Finally, the total expression for $\nu(r)$ is

\begin{equation}
\nu(r) =   \left[ \frac{1} {t_{ff} \times t_{col} } \right]_{gal}^{0.5}
             \times \left( { r_{c} \over r } \right) ^{3 \gamma }
          \times 
\left[\frac{ \overline{\rho}_{g}(r_{c},T_{G}) }
            {\overline{\rho}_{g}(r,T_{G}) }
            \right]^{\gamma} Gyr^{-1}
\label{nu_tot}
\end{equation}

{\it Remark.} Before proceeding further, we like to briefly comment on the 
apparent complexity of the model adopted to perform the analysis. First of 
all, the model has to be considered as gross tool to understand how gradients
in star formation would affect gradients in metallicity and photometric properties. Secondly, the analysis itself is almost model-independent
as what matters here is to make use of a  reasonably grounded
formulation able to predict gradients in chemical abundances and in narrow
band indices and look and the mutual correlation between them. 
\littleskip

{\it Galactic winds.}
The models allow for galactic winds triggered by the energy deposit from Type
I and II supernova  explosions and stellar winds from massive stars. The
formalism in usage here is the same as in Bressan et al. (1994) and Tantalo et
al. (1996) but for two important details: first only a fraction ($\eta=0.3$) of
the kinetic energy by stellar winds is let thermalize, and second the cooling
law for supernova remnants is the same as in Gibson (1994, 1996). When the
total thermal energy of the gas in each region exceeds the local gravitational
potential, gas is supposed to escape from the galaxy and star formation to
stop over there. 

It worth noticing for the sake of clarity that in presence of galactic winds, the
local density  and total  mass of baryonic material in turn can never reach to
the asymptotic values $\rho_L(r,T_G)$ and  $M_{L,12}$, respectively. The
discussion by Tantalo et al. (1997) of this topic clarifies, however, that the
final results of the models are not too severely affected by  contradiction
between the initial hypothesis (models constrained to match the asymptotic
mass) and the real value of this reached in the course of evolution.
\littleskip

{\it Chemical and photometric evolution. } The chemical evolution of elemental species is governed
by the same set of equations as in Tantalo et al. (1996) however adapted to
the density formalism and improved as far as the ejecta and the contribution
from Type Ia and Type II supernovae are concerned according to the revision
made by Portinari et al. (1997) and Tantalo et al. (1997) to whom we refer.
Finally, the line strength indices \Mg2\ and \MFe\ have been calculated
adopting the calibrations by Worthey (1992) and Worthey et al. (1994) as a
function of [Fe/H], $\rm T_{eff}$ and gravity of the stars.
\littleskip

{\it Mass-radius relationship.} The final step is to adopt a  relationship
between $R_{L,e}$ and $M_{L}$  so that assigned the total baryonic mass, the
effective radius and  all other quantities in turn are known. For the purposes
of this study and limited to the case of $H_{0}=50~Km~sec^{-1}Mpc^{-1}$, we
derive  from the data of Carollo et al. (1993) and Goudfrooij et al. (1994)
the following relation

\begin{equation}
       R_{L,e} = 17.13 \times M_{L,12}^{0.557} 
\label{reff_mass}
\end{equation}

\noindent
where $R_{L,e}$ is in kpc. 
\littleskip

{\it Main results.} 
This simple modelling of the distribution of density and hence mass in a
spherical system allow us to describe the gradients in star formation, chemical
abundances, ages, and photometric properties. These   models indeed  are able
to reproduce (i) the slope of colour-magnitude relation by Bower et al.
(1992a,b); (ii) the UV excess as measured by the colour (1550--V) by Burstein
et al. (1988); (iii) the mass to blue luminosity ratio $\rm (M/L_{B})_{\odot}$
of elliptical galaxies. See Tantalo et al. (1997) for all other details.

For the purposes of the discussion below, in Table~1 we present the basic data
for the central core ($ r=0.05 R_{L,e}$) and the first shell ($0.05\times
R_{L,e} \leq r \leq 0.15 \times R_{L,e}$) in a typical galaxy with total
luminous mass $3M_{L,12}$.

The content of Table~1 is as follows: $\Delta M_{L,S}$ is the asymptotic
luminous mass of the region in units of $ 10^{12}$\Msun; $\nu$ is the local
efficiency of the SFR; $\tau$ is the local time scale of gas accretion in Gyr;
$t_{gw}$ is the age in Gyr at which energy heating by  supernova explosions
and stellar winds exceeds the local gravitational energy; $ Z_{max}$ and $
\langle Z \rangle$ are the maximum and mean metallicity, respectively; $ G(t)$
and $ S(t)$ are the local fractionary masses of gas and stars gas,
respectively, both normalized to the asymptotic mass $\Delta M_{L,S}$; $
N_{enh}$ is the percentage of stars showing $\alpha$-enhancement that are
present in the region (see below). Finally, in Fig.~\ref{x_age} we show the
temporal evolution of the abundances of a few elements in the central core of
the model. No detail for the remaining regions is given here but for the
gradients in broad band colors and narrow band indices to be presented in
Fig.~\ref{mod_car} below.

\begin{table}
\begin{center}
\caption{Basic features for the central core and first shell of the reference 
model with $3M_{L,12}$}
\small
\begin{tabular} {l| c | c }
\hline
\hline
 & &  \\
\multicolumn{1}{l|}{Parameter} &
\multicolumn{1}{c|}{ Core} &
\multicolumn{1}{c}{$1^{st}$ shell} \\
 & &  \\
\hline
 & &  \\
 $ \Delta M_{L,S}$       & 0.146  & 0.150  \\
 $\nu$                   & 7.1    & 50.0   \\
 $\tau$                  & 0.74   & 0.29   \\
 $ t_{g\omega}$          & 5.12   & 0.79   \\
 $ Z_{max}$              & 0.0964 & 0.0439 \\
 $ \langle Z \rangle$    & 0.0365 & 0.0286 \\
 $ G(t)$                 & 0.004  & 0.010  \\
 $ S(t)$                 & 0.845  & 0.874  \\
 & & \\
\hline
 & & \\
 $ N_{enh}$           & 45.7\% & 53.9\%  \\
 & & \\
\hline
\hline
\end{tabular}
\end{center}
\label{tab1}
\end{table}

\section{Casting the gradient contradiction}

In Fig.~\ref{mod_car} we compare the theoretical and observational gradients
in broad-band colors (left panel) and line strength indices (right panel) for
our proto-type model. The data are from Carollo \& Danziger (1994a,b).

Remarkably, while the model matches the gradient in broad band colors, it
fails  as far as the  line strength indices \Mg2\ and \MFe\ and their
gradients are concerned. Other cases can be found in the Carollo \& Danziger
(1994a,b) list (they are not shown here for the sake of brevity), in which
either the broad band colors or the line strength  indices are matched, but
the simultaneous fit of the two sets of data is not possible.

The obvious attitude toward these matters would be to ascribe the above
failure to inadequacy of the models (which may certainly be the case) and thus
quit the subject. This  is a point of embarrassment  because there is no
obvious explanation as to why models that successfully reproduce many of the
observed properties of elliptical galaxies (cf. Bressan et al. 1994, 1996;
Tantalo et al. 1996; 1997) fail to match the line strength indices.

In addition to this, and even more relevant to the aims of the present study,
a point of contradiction between chemical structure and line strength indices
is soon evident. The problem is cast as follows. The theoretical gradients
$[\Delta ln Mg_2/ \Delta R] \simeq -0.13$ and $[ \Delta ln \langle Fe \rangle
/ \Delta R] \simeq -0.11$ (with R the galactocentric distance in kpc) are
nearly identical.

\begin{figure*}[]
\psfig{file=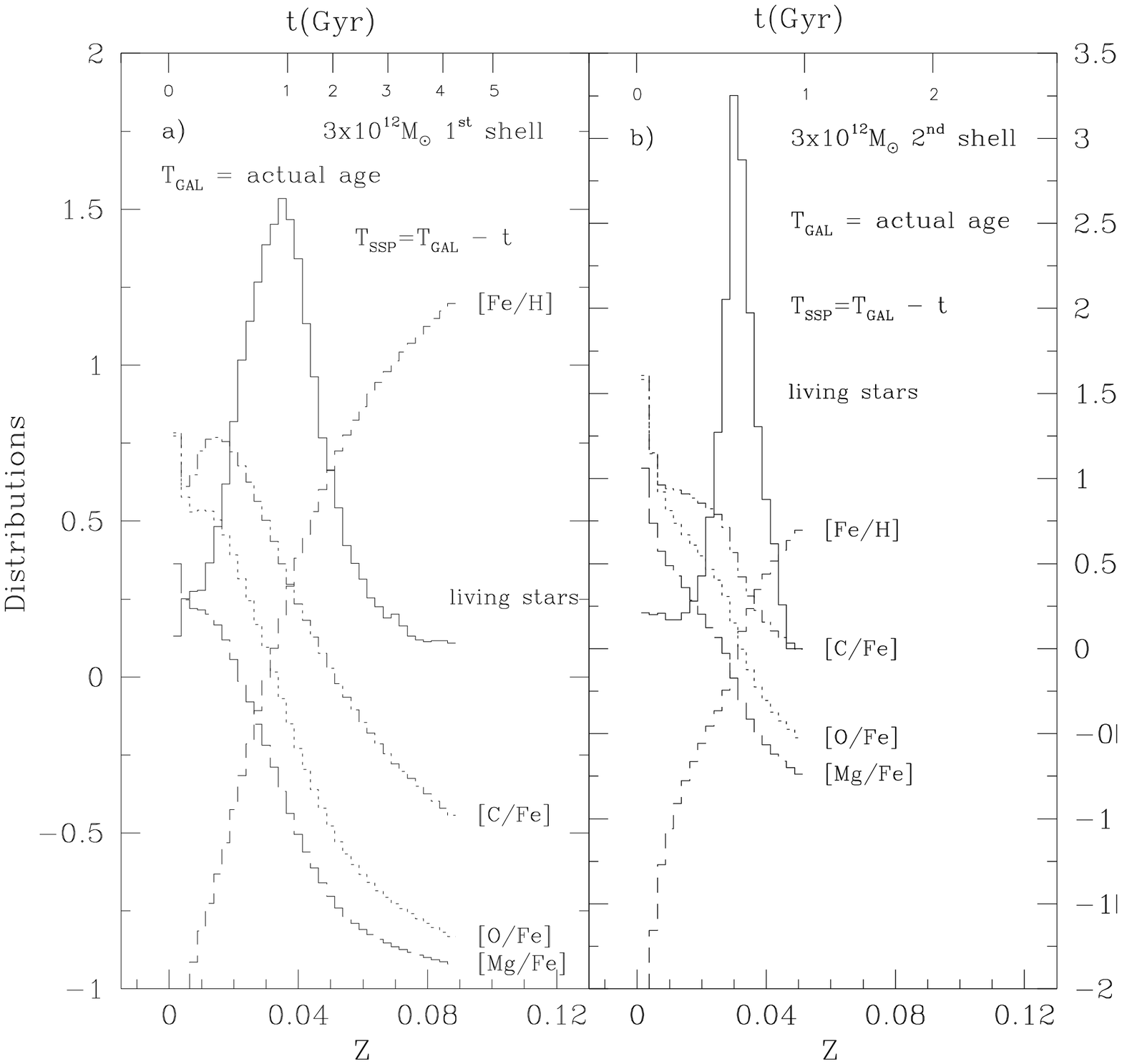,height=9.0truecm,width=17.0truecm} \caption{{\it
Panels (a)} and {\it (b)}: the number of living stars and abundance ratio
distribution per metallicity bin in the central core and the $1^{st}$ shells,
respectively, of the galaxy with mass $ 3M_{L,12}$. The abundance ratios are
in the standard notation. The {\em solid line} is distribution of living stars
in units of $10^{11}M_{\odot}$. The {\em dotted}, {\em dashed}, {\em
long-dashed}, and {\em dot-dashed} lines give the distribution per metallicity
bin of $[O/Fe]$, $[Fe/H]$, $[Mg/Fe]$, and $[C/Fe]$, respectively. The top
scale gives the birth-time $t=T_{G} - T_{SSP}$ in Gyr of a SSP of age
$T_{SSP}$ in a galaxy of age $T_{G}$}
\label{x_nz}
\end{figure*}
 
This is a surprising result because looking at the data of Table~1 the
duration of star formation was much longer in the central core than in the
outer shell (the same trend holds for all remaining shells not considered
here), which implies that the stars in the core are on the average less
$\alpha$-enhanced than in the more external regions (cf. the temporal
evolution of the elemental abundances shown in Fig.~\ref{x_age}), whereas the
gradients we have obtained seem to indicate a nearly constant ratio $[Mg/Fe]$.

To single out  the reason of the contradiction we look at the variation of the
abundance ratios $[Fe/H]$, $[C/Fe]$, $[O/Fe]$, and $[Mg/Fe]$ (with respect to
the solar value) as a function of the metallicity and time, and the present-day
partition  function $N(Z)$. This allows us to evaluate the fraction of living
stars with metallicity above any particular value and with abundance ratios
above or below the solar value. The relationships in question are shown in the
two panels of Fig.~\ref{x_nz} (the left panel is for the central core; the
right panel is for the $1^{st}$ more external shell).

In addition to this, we also look at the current age of the stellar population
stored in every metallicity bin (we remind the reader that the metallicity in
this model in a monotonic increasing function of the age, cf.
Fig.~\ref{x_age}). The top axis of Fig.~\ref{x_nz} shows the correspondence
between metallicity and birth-time of the stellar content in each metallicity
bin, shortly named single stellar population (SSP). The SSP birth-time is $t =
T_{G} - T_{SSP}$, where $T_{G}$ is the present-day galaxy age and $T_{SSP}$ is
the current age of the SSP.
  
From this diagram we learn that the external shell is truly richer in
$\alpha$-enhanced stars ($\sim 53.9\%$ of the total) than the central core
($\sim 45.7\%$ of the total). The percentages $ N_{enh}$ are given in Table~1.
This confirms our expectation that the models should predict gradients in line
strength indices consistent with the gradients in abundances.

{\it Which is the reason for such unexpected contradiction? }

One may argue that the above disagreement results  either from the adoption of
calibrations, such as those by Worthey (1992) and Worthey et al. (1994), which
include the dependence on $[Fe/H]$, $ T_{eff}$, and gravity but neglect the
effect of enhancing the $\alpha$-elements, or the particular model in usage.

\begin{figure}
\psfig{file=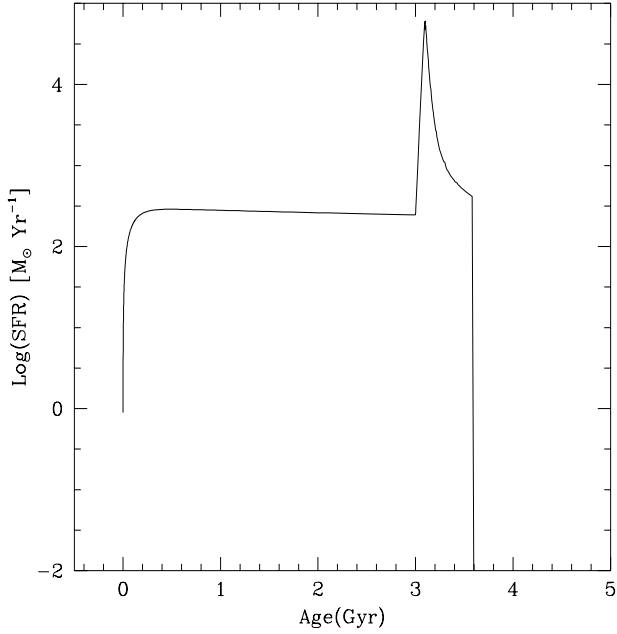,height=9.0truecm,width=8.5truecm} 
\caption{{\it Model-C}: the star formation rate as a function of the time for
the central region of the galaxy model with $3M_{L,12}$. At the age of 3 Gyr
the efficiency of the SFR is let increase  from $\nu = 0.1$ up to $\nu = 50$
over a time scale of  $10^{8}$yr. The parameters of this model are given in
Table~2}
\label{sfr}
\end{figure}

\begin{figure}
\psfig{file=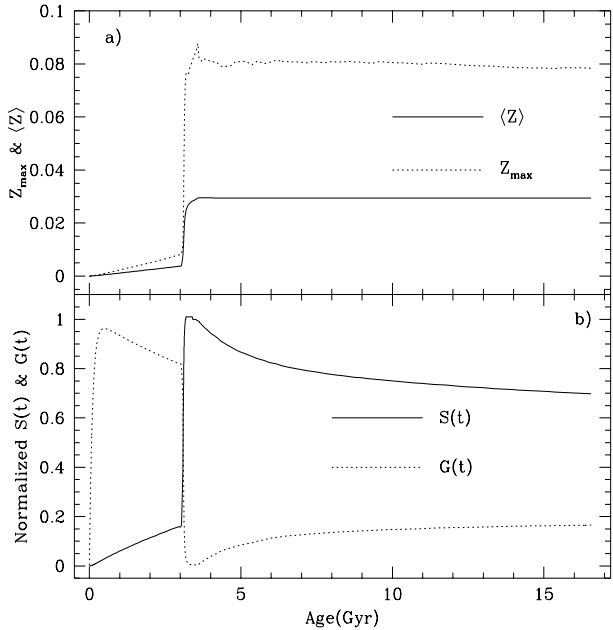,height=9.0truecm,width=8.5truecm}
\caption{{\it Model-C}: Panel (a) shows maximum and mean metallicity {\em
dotted} and {\em solid line}, respectively. Panel (b) shows the fractionary
density of gas $ G(t)$ and living stars $ S(t)$ as a function of time, {\em
dotted} and {\em solid line}, respectively}
\label{chem}
\end{figure}

\begin{table}
\begin{center}
\caption{Basic properties of the central region of the test models with
$3M_{L,12}$. Model-A: late galactic wind and no enhancement of
$\alpha$-elements. Model-B: early galactic wind and enhancement of
$\alpha$-elements. Model-C: recent burst of star formation, and galactic wind,
and strong enhancement of $\alpha$-elements}
\small
\begin{tabular}{l| c | c | c}
\hline
\hline
 & &  \\
\multicolumn{1}{l|}{Parameter} &
\multicolumn{1}{c|}{Model-A} &
\multicolumn{1}{c|}{Model-B} &
\multicolumn{1}{c}{Model-C} \\
 & & & \\
\hline
 & & & \\
 $ \Delta M_{L,S}$       & 0.146  & 0.146  & 0.146         \\
 $\nu$                   & 7.1    & 100.0  & 0.1 $\div$ 50 \\
 $\tau$                  & 0.74   & 0.05   & 0.10          \\
 $t_{g\omega}$           & 5.12   & 0.39   & 3.58          \\
 $Z_{max}$               & 0.0964 & 0.0713 & 0.0878        \\
 $\langle Z \rangle$     & 0.0365 & 0.0279 & 0.0294        \\
 $G(t)$                  & 0.004  & 0.002  & 0.003         \\
 $S(t)$                  & 0.845  & 0.942  & 0.994         \\
 & & & \\
\hline
 & & & \\
 $N_{enh}$               & 45.7\% & 85.2\% & 74.8\% \\
 & & & \\
\hline
\hline
\end{tabular}
\end{center}
\label{tab2}
\end{table}

\section{Changing calibrations and chemistry }

To answer the question posed in the previous section, first we adopt a
different calibration in which the effect of $[Mg/Fe]$ is explicitly taken
into account, and second we discuss different, {\it ad hoc} designed, galactic
models in which different levels of enhancement in \alfa\ are let occur by
artificially changing the history of star formation.

\subsection{Calibrations containing [Mg/Fe]}

Many studies have emphasized that line strength indices depend not only on the
stellar parameters $T_{eff}$ and gravity, but also on the chemical abundances
(Barbuy 1994, Idiart et al. 1995, Weiss et al. 1995, Borges et al. 1995).

We start pointing out that in presence of a certain degree of enhancement in
$\alpha$-elements one has to suitably modify relationship between the total
metallicity $Z$ and the iron content $[Fe/H]$. Using the pattern of abundances
by Anders \& Grevesse (1989), Grevesse (1991) and Grevesse \& Noels (1993), we
find the  general  relation

\begin{equation}
\left[\frac{Fe}{H} \right] = \log{\left(\frac{Z}{Z_{\odot}}\right)} -
 \log{\left(\frac{X}{X_{\odot}}\right)} -
0.8\left[\frac{\alpha}{Fe}\right] - 0.05\left[\frac{\alpha}{Fe} \right]^{2}
\label{feh}
\end{equation}

\noindent 
where the term $[\alpha/Fe]$ stand for all $\alpha$-elements lumped together.
 
The recent empirical calibration by Borges et al. (1995) for the \Mg2\ index
includes the effect of different $[Mg/Fe]$ ratios

\begin{displaymath}
 {\ln}{Mg_{2}} = -9.037 + 5.795 \frac{5040}{T_{eff}} + 0.398 \log{g}
+ ~~~~~~~~
\end{displaymath}
\begin{equation}
 ~~~~~~~~0.389 \left[ \frac{Fe}{H} \right] - 0.16 \left[ \frac{Fe}{H}
\right]^{2} + 0.981 \left[ \frac{Mg}{Fe} \right] 
\label{mg2}
\end{equation}

\noindent
which holds for effective temperatures and gravities in the ranges $ 3800  <
T_{eff} < 6500$ K and $ 0.7 < \log{g} < 4.5$.

To our knowledge, no corresponding calibration for the \MFe\  index is yet
available, so that one is forced to adopt that with no dependence on
$[Mg/Fe]$. Nevertheless, a zero-order evaluation of the effect of $[Mg/Fe]$ on
the \MFe\ index is possible via the different relation between $Z$ and
$[Fe/H]$ of the $[\alpha/Fe] \neq 0$ case.

The above relations for $[Fe/H]$, \Mg2\ and implicitly \MFe\ are used to
generate new SSPs and galactic models in which not only the chemical
abundances are enhanced with respect to the solar value but also the effect of
this on the line strength indices is taken into account in a self-consistent
manner.

\begin{figure*}[]
\psfig{file=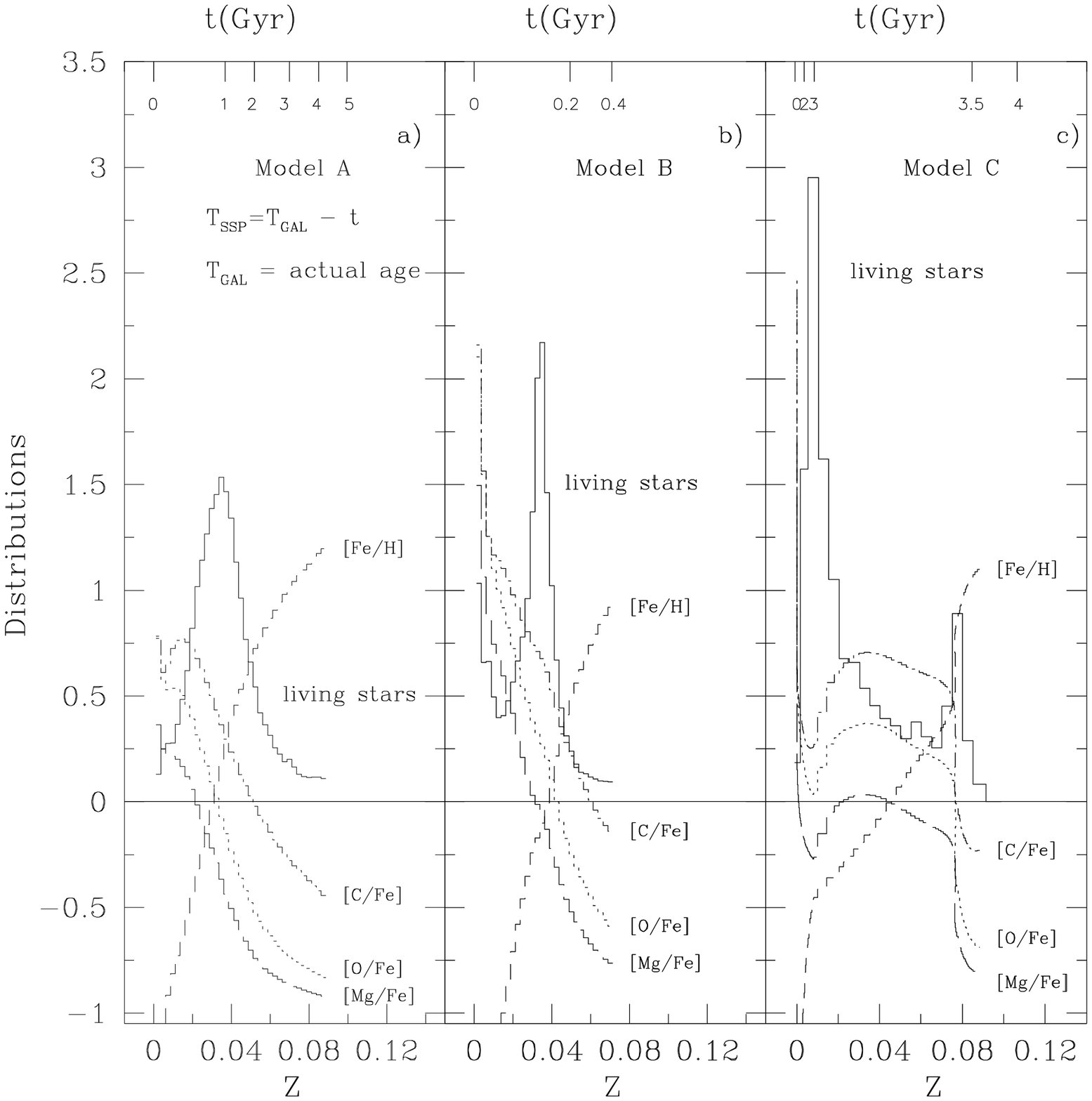,height=9.0truecm,width=17.0truecm}
\caption{The partition function $N(Z)$ and abundance ratios distribution per
metallicity bin, for the Model-A (left panel), Model-B (middle panel), and
Model-C (right panel). The {\em solid line} is $N(Z)$ in units of $10^{11}$ at
the age of 15 Gyr. The {\em dotted}, {\em dashed}, {\em long-dashed}, and {\em
dot-dashed} lines give the distribution per metallicity bin for $[O/Fe]$,
$[Fe/H]$, $[Mg/Fe]$ and $[C/Fe]$, respectively. The abundance ratios are in the
standard notation. The top scale gives the birth-time $t=T_{G}-T_{SSP}$ of a
SSP with age $T_{SSP}$ in a galaxy with age $T_G$}
\label{x_nz_abc}
\end{figure*}

\subsection{Three different chemical structures }

We present here three galactic models that in virtue of their particular
history of star formation, have different chemical structures and degree of
enhancement in \alfa. The discussion is limited to the central region of the
$3M_{L,12}$ galaxy.
\littleskip
  
{\it Model-A: late galactic wind}.
This case has late galactic wind ($\sim 5.12$ Gyr), which means that  Type Ia
supernovae dominate the enrichment in $Fe$ of the gas, and the ratio
[$\alpha/Fe$] is solar or below solar for most of the time. This model is
actually the central region of the $3M_{L,12}$ galaxy presented above. The
percentage ($ N_{enh}$) of $\alpha$-enhanced stars that are still alive at the
age of 15 Gyr amounts to 45.7\%.
\littleskip

{\it Model-B: early galactic wind}.
In order to enhance the relative abundance of elements from Type II supernovae
we arbitrarily shortened the duration of the star forming period. To this aim,
in the central region of the same galaxy, the efficiency of star formation has
been increased $\nu=100$) and the infall time scale decreases ($\tau$=0.05
Gyr) so that the galactic wind occurs much earlier  than in the previous case
(at 0.39 Gyr). The material (gas and stars) of Model-B is therefore strongly
enhanced in \alfa. The percentage ($ N_{enh}$) of $\alpha$-enhanced stars that
are still alive at the age of 15 Gyr amounts to 87.2\%.
\littleskip

{\it Model-C: recent burst of star formation}. 
A third possibility is considered, in which a burst of star formation can
occur within a galaxy that already underwent significant stellar activity and
metal enrichment during its previous history. This model (always limited to
the central region of the galaxy) has nearly constant star formation rate from
the beginning, but at the age (arbitrarily chosen) of 3 Gyr it is supposed to
undergo a sudden increase in the star formation rate. To this aim, the
specific efficiency of star formation $\nu$ is let increase from $\nu=0.1$ to
$\nu=50$ over a time scale of $10^{8}$~yr. The initial nearly constant stellar
activity is secured by adopting a long time scale of gas accretion in the
infall scheme ($\tau=10$ Gyr). The rate of star formation (in units of
\Msun$\rm yr^{-1}$)  as a function of time  is shown in Fig.~\ref{sfr}. Soon
after the intense period of star formation, the galactic wind occurs thus
halting  star formation and chemical enrichment. The basic chemical properties
of the model as a function of the age are shown in Fig.~\ref{chem}. This
displays the maximum ({\em dotted line}) and mean metallicity ({\em solid
line}), and the fractionary mass of gas $ G(t)$ ({\em dotted line}) and stars
$ S(t)$ ({\em solid line}) both normalized to $\Delta M_{L,S}$. The percentage
($ N_{enh}$) of $\alpha$-enhanced stars that are still alive at the age of 15
Gyr amounts to 74.5\%.

The basic data for the three models in question are summarized in Table~2,
whereas the evolution of their  chemical abundances and  present-day partition
function $N(Z)$ are  shown in Figs.~\ref{x_nz_abc}, where the left panel is for
Model-A, the middle panel is for Model-B, and the  right panel is for Model-C.
These figures are the analog of Fig.~\ref{x_nz}.

\subsection{\Mg2\ and \MFe\ in SSPs with $[\alpha/Fe]\neq 0$ } 

To secure full consistency between chemical abundances and line strength
indices for  the galactic models and make use of the SSP technique (cf.
Bressan et al. 1996) one should adopt SSPs with the same pattern of abundances
indicated by the chemical models.

To this aim, the distributions of chemical abundances (and their ratios) as a
function of the total metallicity and/or time  provided by the model galaxies
are used  to calculate the \Mg2\ and \MFe\ indices of SSPs with different $Z$,
$[Fe/H]$, $[O/Fe]$, and $[Mg/Fe]$.

\begin{figure}
\psfig{file=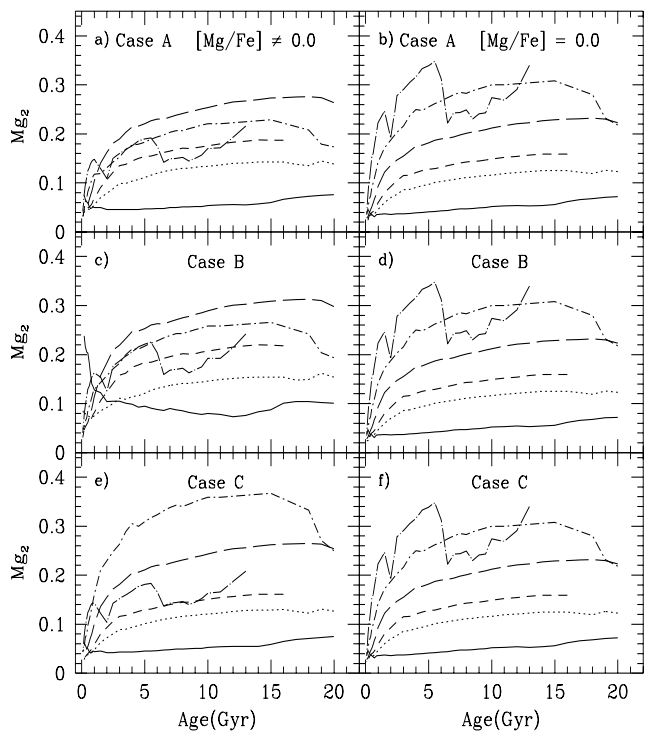,height=10.5truecm,width=9.0truecm}
\caption{{\it Panels (a)}, {\it (c)} and {\it e)} show the \Mg2\ index
evolution for SSPs with different metallicity (Z=0.0004, Z=0.004, Z=0.008,
Z=0.02, Z=0.05 and Z=0.1 {\it solid} {\it dotted} {\it dashed} {\it
long-dashed}, {\it dot-dashed} and {\it dot-long-dashed} lines, respectively)
and the assumption of enhancement in $\alpha$-elements. {\it Panels (b)}, {\it
(d)} and {\it (f)} show the same but without enhancement of $\alpha$-elements}
\label{mg2_ssp}
\end{figure}

\begin{figure}
\psfig{file=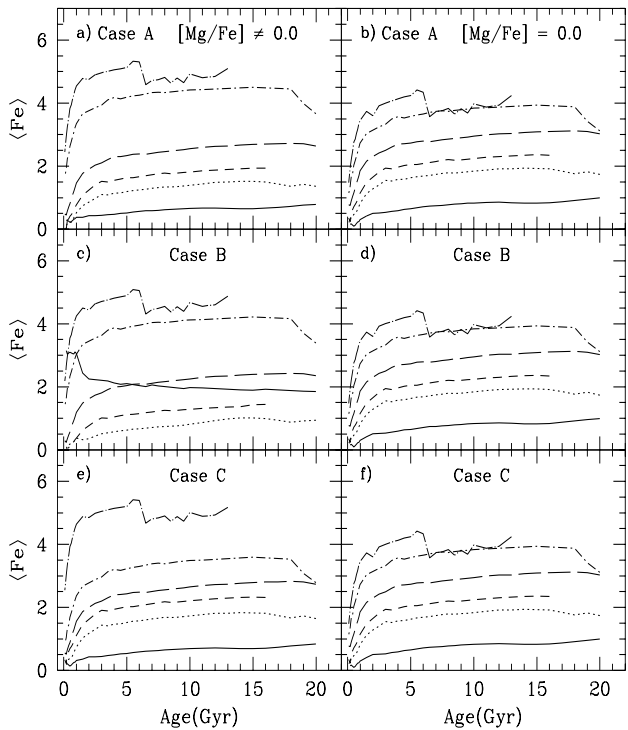,height=10.5truecm,width=9.0truecm}
\caption{{\it Panels (a)}, {\it (c)} and {\it e)} show the \MFe\ index
evolution for SSPs with different metallicity (Z=0.0004, Z=0.004, Z=0.008,
Z=0.02, Z=0.05 and Z=0.1 {\it solid} {\it dotted} {\it dashed} {\it
long-dashed}, {\it dot-dashed} and {\it dot-long-dashed} lines, respectively)
and the assumption of enhancement in $\alpha$-elements. {\it Panels (b)}, {\it
(d)} and {\it (f)} show the same but without enhancement of $\alpha$-elements}
\label{fe_ssp}
\end{figure}

To quantify the abundance of $\alpha$-elements we prefer to use $[O/Fe]$
instead of  $[Mg/Fe]$, because the stellar yields by Portinari et al. (1997)
that are at the base of our chemical models somewhat overestimate the
production of $Fe$ by Type II supernovae, and underestimate in turn the ratio
$[Mg/Fe]$ as compared to the observational value. According to Portinari et
al. (1997) the theoretical $[Mg/Fe]$ is about $0.2\div 0.3$ dex lower than
indicated by the observational data. Applying this correction, the ratio
$[Mg/Fe]$ gets close to the ratio $[O/Fe]$ (see Fig.~\ref{x_nz_abc}), which
somehow justifies our use of $[O/Fe]$ instead of $[Mg/Fe]$ in the procedure
below. This marginal drawback of the chemical model does not however affect
the conclusion of this analysis.

For any value of $Z$ we read from Fig.~\ref{x_nz_abc} the ratios $[C/Fe]$,
$[O/Fe]$, and $[Mg/Fe]$, derive $[Fe/H]$ from equation (\ref{feh}), and insert
$[Mg/Fe], i.e. =[O/Fe]$, and $[Fe/H]$ into equation (\ref{mg2}). It goes
without saying that the values of $[Fe/H]$  derived from Figs.~\ref{x_nz_abc}
and equation (\ref{feh}) are mutually consistent by definition.

Table~3 shows the values of $[Fe/H]$ and $[O/Fe]$ assigned to the SSPs
according to the chemical structure of Model-A, Model-B, and Model-C and, for
purposes of comparison, to  reference SSPs with no enhancement at all.

The temporal evolution of the \Mg2\ and \MFe\ indices for the SSPs listed in
Table~3 is shown in the various panels of Figs.~\ref{mg2_ssp} and
~\ref{fe_ssp}, respectively.

\begin{figure*}[t]
\psfig{file=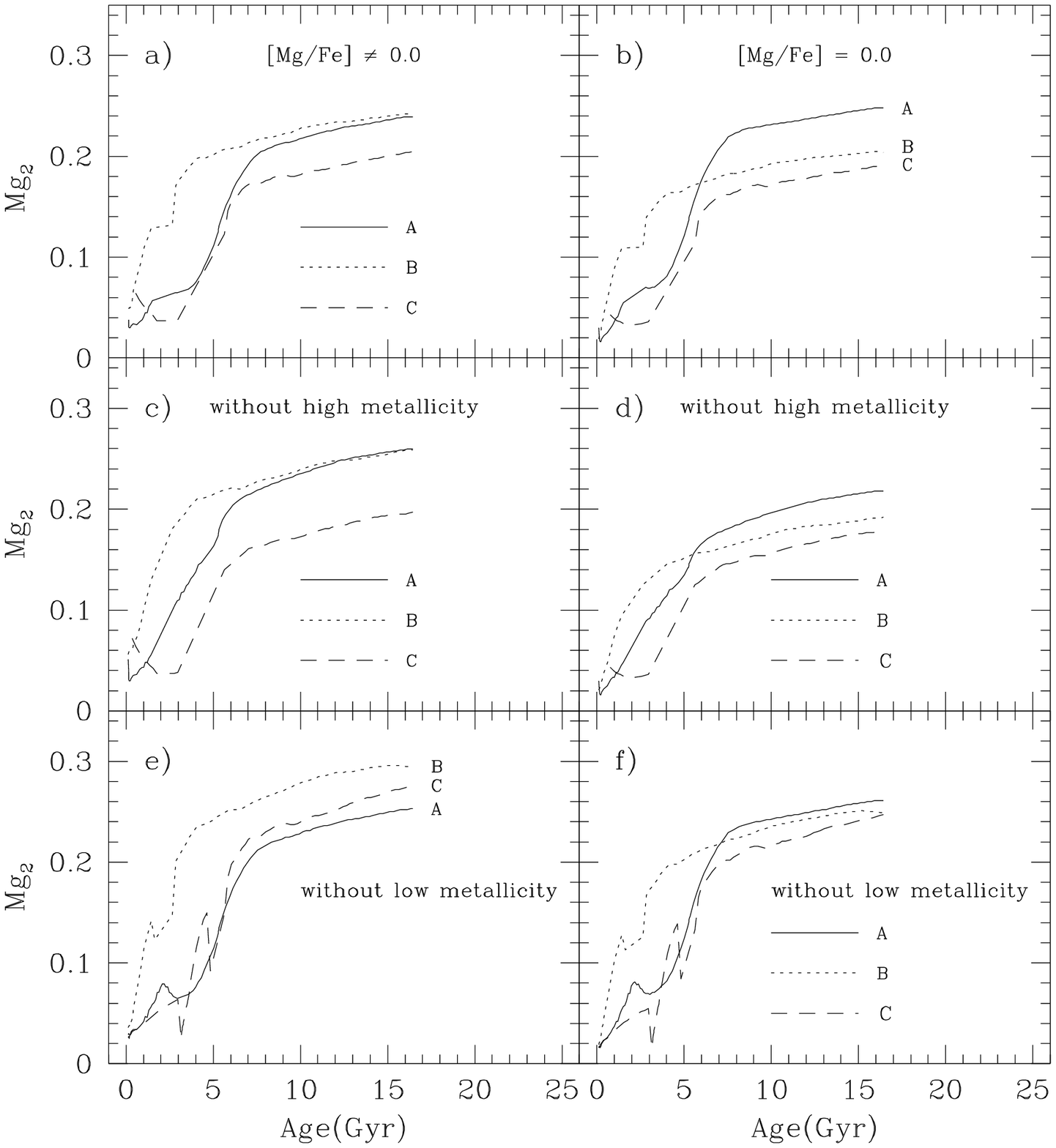,height=9.0truecm,width=17.0truecm}
\caption{Evolution of the \Mg2\ index as a function of time. {\it Panels (a)},
{\it (c)} and {\it (e)} show the evolution of the \Mg2\ index calculated
including the effect of the chemical abundances, while {\it Panels (b)}, {\it
(d)} and {\it (f)} show the same but without the effect of
$\alpha$-enhancement. The {\em solid line} corresponds to Model-A the {\em
dotted line} to Model-B, and {\em dashed line} to Model-C}
\label{mg2_abc}
\end{figure*}

\begin{figure*}[t]
\psfig{file=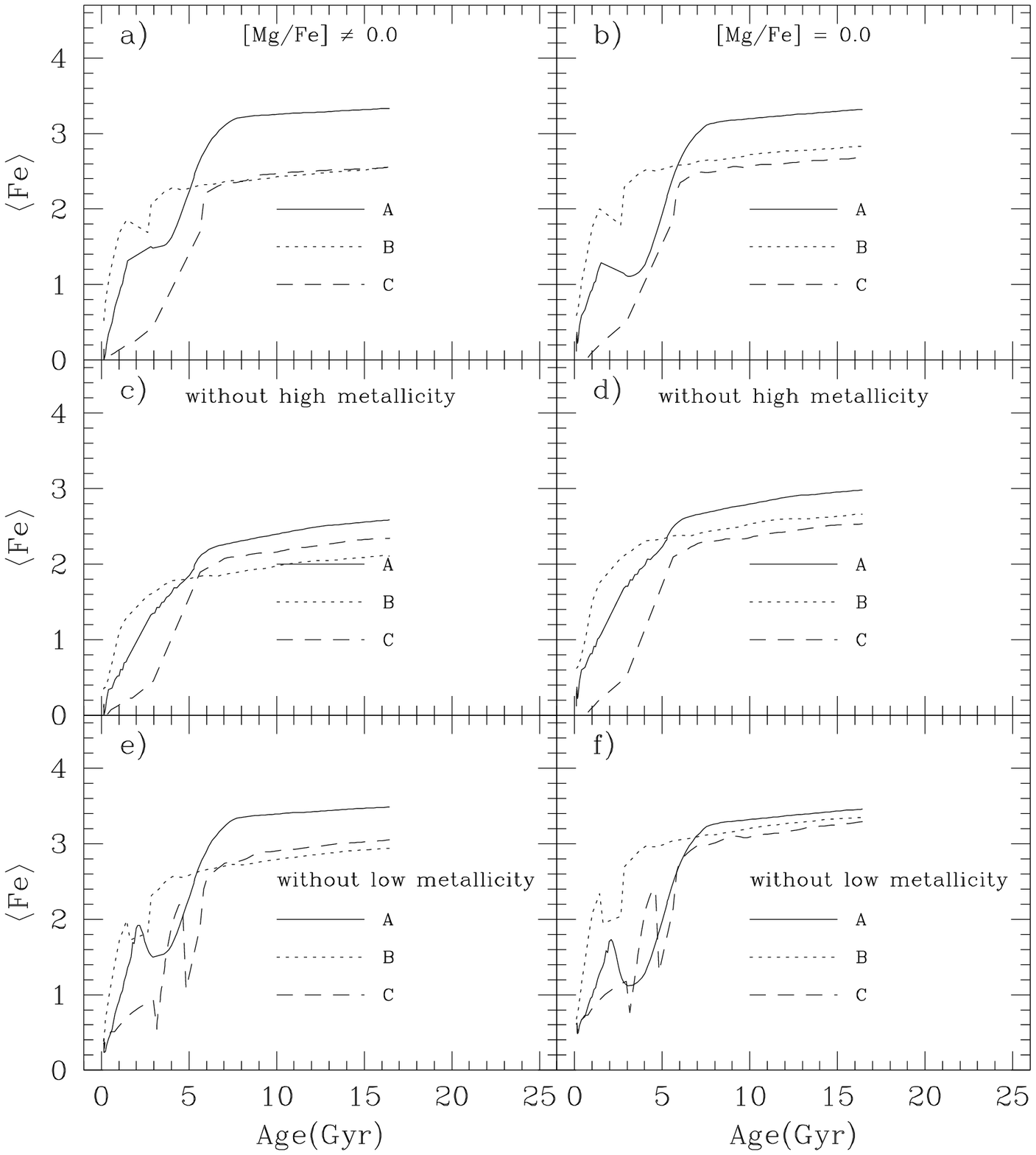,height=10.5truecm,width=17.0truecm}
\caption{Evolution of the \MFe\ index as a function of time. {\it Panels (a)},
{\it (c)} and {\it (e)} show the evolution of the \Mg2\ index calculated
including the effect of the chemical abundances, while {\it Panels (b)}, {\it
(d)} and {\it (f)} show the same but without the effect of
$\alpha$-enhancement. The {\em solid line} corresponds to Model-A the {\em
dotted line} to Model-B and {\em dashed line} to Model-C}
\label{fe_abc}
\end{figure*}

\begin{table*}
\begin{center}
\caption{$[O/Fe]$ and $[Fe/H]$ ratios for SSPs with enhancement of \alfa\
according to Model-A, Model-B and Model-C. The same ratios for the reference
SSPs with no enhancement are also shown}
\scriptsize
\begin{tabular*}{113mm} {l| c c| c c| c c| c c}
\hline
\hline
 & & & & & & & & \\
\multicolumn{1}{l|}{Z} &
\multicolumn{2}{c|}{Model-A} &
\multicolumn{2}{c|}{Model-B} &
\multicolumn{2}{c|}{Model-C} &
\multicolumn{2}{c}{Reference SSPs} \\
\hline
 & & & & & & & & \\
 & $[O/Fe]$ & $[Fe/H]$ & $[O/Fe]$ & $[Fe/H]$ & $[O/Fe]$ & $[Fe/H]$ 
        & $[O/Fe]$ & $[Fe/H]$ \\
 & & & & & & & & \\
\hline
 & & & & & & & & \\
 0.0004 & +0.8   & --2.38 & +3.28  & --4.87 & +0.51   & --2.13 & 0.0 & --1.71 \\
 0.004  & +0.6   & --1.22 & +1.72  & --2.23 & +0.14   & --0.82 & 0.0 & --0.71 \\
 0.008  & +0.5   & --0.80 & +1.22  & --1.44 & +0.03   & --0.42 & 0.0 & --0.39 \\
 0.02   & +0.4   & --0.30 & +0.72  & --0.57 & +0.30   & --0.22 & 0.0 &  +0.03 \\
 0.05   & --0.50 & +0.88  & --0.25 & +0.69  & +0.31   & +0.24  & 0.0 &  +0.50 \\
 0.1    & --0.80 & +1.55  & --0.59 & +1.40  & --0.87  & +1.60  & 0.0 &  +0.95 \\
 & & & & & & & & \\
\hline
\hline
\end{tabular*}
\end{center}
\label{tab3}
\end{table*}

It is soon evident that with $[\alpha/Fe]=0$ (right panels of
Fig.~\ref{mg2_ssp}), \Mg2\ monotonically increases with the metallicity, but
for the extreme SSP with Z=0.1 for which the trend is reversed at ages older
than 5 Gyr. When $[\alpha/Fe]\neq 0$, the trend is more complicated as it
depends on the degree of enhancement. For cases A and B, the strongest \Mg2\
happens to occur for $Z=0.02$, whereas for case C it occurs for $Z=0.05$ (left
panels of Fig.~\ref{mg2_ssp}).

As far as \MFe\ is concerned, with $[\alpha/Fe]=0$ the index gets stronger at
increasing metallicity but for the extreme case of $Z=0.1$, in which \MFe\
gets lower than or comparable to the values for the case with $Z=0.05$  at
ages older than about 5 Gyr (right panels of Fig.~\ref{fe_ssp}). In presence
of enhancement in \alfa, there is a significant dependence of \MFe\ on this
parameter even at our zero-order evaluation (see the left panels of
Fig.~\ref{fe_ssp}). Although the exact evaluation of the effect of $[Mg/Fe]$
on \MFe\ is hampered by the lack of the proper calibration, still the above
experiments clarify that it cannot be neglected.

\subsection { The \Mg2\ and \MFe\ indices of galaxies}

{\it What the results for the \Mg2\ and \MFe\ indices would be when applying
these SSPs  to model galaxies ? } The situation is displayed in the various
panels of Fig.~\ref{mg2_abc} for \Mg2\ and Fig.~\ref{fe_abc} for \MFe. The
combined analysis of the chemical structures, partition functions $N(Z)$, and
temporal variations of the \Mg2\ and \MFe\ indices of the model galaxies
(Figs.~\ref{mg2_abc} and ~\ref{fe_abc}) allow us to make  the following
remarks:

\begin{description}

\item [(i)] $[Mg/Fe]\neq 0$: \Mg2\ in Model-A (late wind, no chemical
enhancement) is always weaker than in Model-B (early wind, significant
chemical enhancement). However, the difference is large for ages younger than
about 5 Gyr, and gets very small up to vanishing for older ages. \Mg2\ of
Model-C is always weaker than Model-A and Model-B which means that the higher
(or comparable) enhancement in \alfa\ of Model-C with respect to the previous
ones does not produce a stronger \Mg2\ index. The age dependence of \MFe\ is
more intrigued. At young ages ($< 5$ Gyr) the intensity of \MFe\ gets stronger
passing from Model-C to Model-A and finally Model-B. At ages older than 5 Gyr,
Model-A has the strongest \MFe\ whereas Model-B and Model-C are weaker and of
the same intensity.

\item [(ii)] $[Mg/Fe] = 0$: \Mg2\ in Model-A (late wind, no chemical
enhancement) is first weaker than in Model-B (early wind, significant chemical
enhancement) up to ages of about 5 Gyr, and then becomes significantly
stronger at older ages.  Finally, the \Mg2\ index of Model-C (burst of star
formation, and strong enhancement) is weaker or about equal to that of Model-A
and Model-B. The index \MFe\  closely follows the trend of \Mg2\ all over the
age range.
\end{description}

It follows that both \Mg2\ and \MFe\ don not simply correlate with age,
$[Fe/H]$ and $[Mg/Fe]$. The striking result is that \Mg2\ and to some extent
\MFe\ as well of models with supposedly the highest degree of enhancement in
\alfa\ happen to be weaker than in those with no enhancement.

{\it What causes the odd behaviour of  \Mg2\ and \MFe\  as a function of the
age and underlying chemical structure of the model galaxy ? }
\littleskip

To answer this question, we have artificially removed from the partition
function $N(Z)$ all the stars in certain metallicity bins and re-calculated the
line strength indices for the three models.

Panels (c) and (d) of  Figs.~\ref{mg2_abc} and ~\ref{fe_abc} show the results
when all stars with metallicity higher than $Z$=0.05 are removed. This is
motivated by the trend as function of the metallicity shown by the SSPs we
have already pointed out. The situation remains practically unchanged.

Likewise, panels (e) and (f) of Figs.~\ref{mg2_abc} and ~\ref{fe_abc}  show the
same but when all stars with metallicity lower than $Z=0.008$ are removed. Now
the results change significantly. In the case of $[Mg/Fe] \neq 0$, the \Mg2\
index of Model-B is always much stronger than that of Model-A. In such a case
there is correspondence between the strength of the index and the amount of
enhancement in \alfa. In contrast, \MFe\ of Model-A is first slightly weaker
and then stronger than in Model-B, whereas that of Model-C is first weaker and
then almost equal to that of Model-B. cases. In the case of  $[Mg/Fe]=0$, at
ages older than about 5 Gyr the \Mg2\ and \MFe\ of the three models are almost
equal each other, with a marginal increase passing from Model-C to Model-B and
Model-A. At younger ages, while Model-A and Model-C have nearly the same
indices, Model-B has always the strongest values.

The above analysis clarifies in a quantitative fashion a number of important
clues:
 
\begin{itemize}
\item{In galaxies, both  \Mg2\ and \MFe\  depend on the age, $N(Z)$, $[Fe/H]$,
and $[Mg/Fe]$ in a somewhat unpredictable fashion.}

\item {Because of this, inferring from the above indices the abundance ratio
$[Mg/Fe]$ and the enhancement of $\alpha$-elements is  a difficult task with
somewhat ambiguous results.}

\item{Different slopes of  the gradients in \Mg2\ and \MFe\ do not
automatically imply gradients in chemical abundances or enhancement ratios.}
\end{itemize}

\begin{figure}
\psfig{file=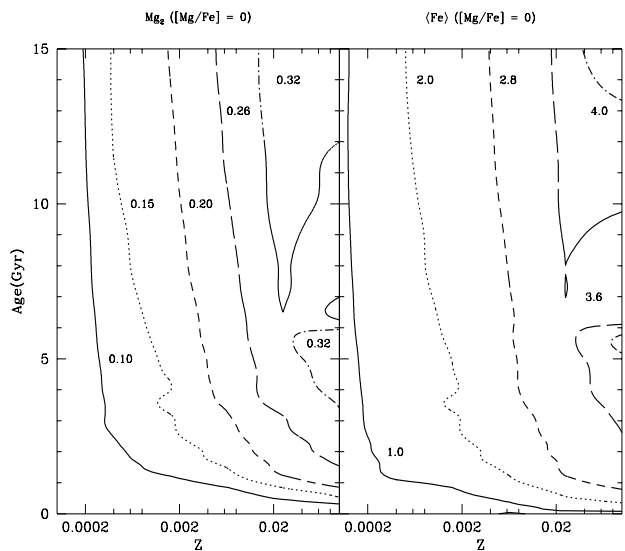,height=9.0truecm,width=8.5truecm}
\caption{ Curves of constant \Mg2\ (left) and \MFe\ (right) at varying age (in
Gyr) and metallicity (Z) for SSP with no enhancement  in \alfa. The
calibrations in use are from Worthey et al. (1994). The intensity of the
indices is annotated along the curves }
\label{level_noenh}
\end{figure}

\begin{figure}
\psfig{file=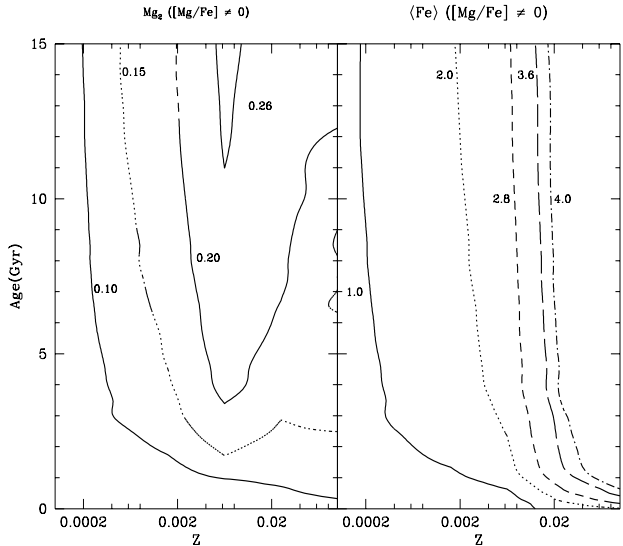,height=9.0truecm,width=8.5truecm}
\caption{ The same as in Fig.~11 but for SSP with enhancement if \alfa\
according to the entries of Table~3. The calibrations in use is from Borges et
al. (1995) for \Mg2\ and Worthey et al. (1994) for \MFe\ however corrected for
the different relation between $Z$ and $[Fe/H]$ in presence of enhancement in
\alfa. The intensity of the indices is annotated along the curves }
\label{level_enh}
\end{figure}

\section{Indices in the age-metallicity plane of SSPs}

We like to conclude this study presenting the loci of  constant \Mg2\ and
\MFe\ in the age-metallicity plane of SSPs. This is shown in 
Figs.~\ref{level_noenh} and \ref{level_enh} displaying the above indices with
and without enhancement of \alfa\ using the calibrations discussed in the
previous sections.

This plane can be used to quickly check how gradients in ages and/or
metallicities across galaxies (inferred from other independent analysis) would
reflect onto gradients in \Mg2\ and \MFe\  and vice-versa.

An interesting feature to note, is the marked dip in   the \Mg2\ index for
values greater than a certain limit that occurs at a certain value of the
metallicity. The threshold value for  \Mg2\ is about 0.26 and the metallicity
is about 0.03 in the case of no enhancement,
 whereas they are  lowered  to
0.15 and 0.008, respectively,  in presence of enhancement.

Starting from the observational result that in general elliptical galaxies
show stronger \Mg2\ and \MFe\ toward the center, and representing the local
mix of stellar population in a galaxy (center and/or external regions) with a
mean SSP of suitable age and composition, we see that the observational
gradient in  \Mg2\ and \MFe\ could be compatible with (i) either a nucleus
more metal-rich and older than the external regions; (ii) or nucleus more
metal-rich and younger than the external regions. In contrast a nucleus  less
metal-rich and older than the external regions would lead to  gradients in
\Mg2\ and \MFe\ opposite to what observed. The situation is straightforward
with the Worthey (1992) calibrations, whereas it is somewhat intrigued with
the Borges et al. (1995) calibration in presence of $\alpha$-enhancement. It
worth recalling here that Bressan et al. (1996) analyzing the Gonzales (1993)
\Hbeta\ and \MgFe\ data for elliptical galaxies and their variation across
these systems suggested that most galaxies ought to have nuclei with higher
metallicities and longer durations of star forming activity than the
peripheral regions. This suggestion is fully compatible with the gradients in
\Mg2\ and \MFe\ observed in elliptical galaxies.

\section{Summary and conclusions}

Aim of this study was to ascertain on a quantitative basis the effect of age,
metallicity, partition function $N(Z)$, abundance ratio $[Mg/Fe]$,  and
calibration in usage on the line strength indices \Mg2\ and \MFe\, from whose
gradients the problem of the possible enhancement of $\alpha$-elements toward
the center of elliptical galaxies originates. Although by no means we want to
exclude such a possibility, the attention is called on a number of indirect
effects that could invalidate the one-to-one correlation between  the index
intensity  and the abundance of the corresponding element, and indirectly on
the one-to-one correlation between the relative slopes of the observational
gradients and the inferred spatial variation of abundance ratios
([$\alpha/Fe]$ in particular).

The results of this study can be summarized as follows:
 
\begin{enumerate}
\item{The intensity of \Mg2\ does not simply correlate with the abundance of
$Mg$ and the ratio $[Mg/Fe]$ in particular.}

\item{The intensity of \Mg2\ does not simply correlate with the age or the
metallicity.}

\item{The intensity of \Mg2\ much depends on the partition function $N(Z)$.}

\item{Likewise for the \MFe\ index.}

\item{ Inferring the abundance of $Mg$ or the enhancement ratio $[Mg/Fe]$ is a
cumbersome affair whose solution is not always possible because hints on $N(Z)$
are needed.}

\item{The observational gradients in \Mg2\ and \MFe\ do not automatically imply
gradients in the abundances of $Mg$ and $Fe$ or enhancement ratios. Inferring
from the observational \Mg2\ and \MFe\ constraints on the past history of star
formation (via the different time scales of $Mg$ and $Fe$ enrichment) may be
risky.}
\end{enumerate}

Although most of these conclusions have already been around in literature,
their quantitative assessment has never been attempted before in a systematic
fashion, in particular in the complex but realistic situation in which many
stellar populations with different ages, metallicities, and abundance ratios
are present.
\littleskip

\acknowledgements{We are most grateful to Dr. Guy Worthey for his constructive referee
report which much contributed to improve upon the orginal manuscript. This study has been financed by the Italian Ministry of
University, Scientific Research and Technology (MURST), the Italian Space
Agency (ASI), and the European Community TMR grant \#ERBFMRX-CT96-0086. }

\newpage
\newpage

\end{document}